\documentclass[11pt]{article}

\usepackage[usenames,dvipsnames,svgnames,table]{xcolor}
\ifdefined\directlua
  \usepackage{fontspec}
\else
  \usepackage[T1]{fontenc}
  \usepackage[nomath]{lmodern}
\fi
\usepackage{arxiv}
\usepackage{booktabs}
\usepackage{graphics, float, url}
\usepackage{graphicx}
\usepackage{subcaption}
\usepackage{hyperref}
\hypersetup{hidelinks}

\usepackage{nicefrac}       
\usepackage{microtype}      

\usepackage{makecell}
\usepackage{multirow}
\usepackage{stmaryrd}
\usepackage{hhline}
\usepackage{fontawesome}
\usepackage{dsfont}
\usepackage{lscape}
\usepackage{colortbl}
\usepackage{bbding}
\usepackage{amsfonts}
\usepackage{mathtools}
\usepackage{amsmath}
\usepackage{amssymb}

\usepackage{array}
\usepackage{url}
\usepackage{tablefootnote}
\usepackage{amssymb}
\usepackage{xcolor}
\usepackage{multicol}
\usepackage{todonotes, marginnote}
\usepackage{changepage,threeparttable} 
\usepackage{booktabs, multirow} 
\usepackage{soul}

\let\tmp\oddsidemargin
\let\oddsidemargin\evensidemargin
\let\evensidemargin\tmp
\reversemarginpar

\input{font_config.sty}

\input{paper_config.tex}
\usepackage{lineno}
\hypersetup{colorlinks,allcolors=black}

\usepackage{tikz}
\usetikzlibrary{positioning}
\usetikzlibrary[shapes.geometric]
\usetikzlibrary{arrows.meta}

\usepackage{marvosym}

\modulolinenumbers[5]
\newcolumntype{L}[1]{>{\raggedright\let\newline\\\arraybackslash\hspace{0pt}}m{#1}}
\newcolumntype{C}[1]{>{\centering\let\newline\\\arraybackslash\hspace{0pt}}m{#1}}
\newcolumntype{R}[1]{>{\raggedleft\let\newline\\\arraybackslash\hspace{0pt}}m{#1}}

\newcommand{\context}[1]{\textcolor{black}{#1}}
\newcommand{\challenge}[1]{\textcolor{black}{#1}}
\newcommand{\hypothesis}[1]{\textcolor{black}{#1}}
\newcommand{\proposal}[1]{\textcolor{black}{#1}}
\newcommand{\evaluation}[1]{\textcolor{black}{#1}}
\newcommand{\paperdesc}[1]{\textcolor{black}{#1}}
\newcommand{\paco}[1]{{\textcolor{black}{#1}}}
\newcommand{\rab}[1]{{\textcolor{black}{#1}}}
\newcommand{\final}[1]{{\textcolor{black}{#1}}}
\newcommand{\major}[1]{{\textcolor{black}{#1}}}

\usepackage[inline]{enumitem} 
\usepackage{graphicx}
\usepackage{verbatim}
\usetikzlibrary{mindmap,shadows}
\usepackage{tabularx}

\title{\mytitle}
\date{}

\author{ 
	Nuria Rodríguez-Barroso $^{*\text{,a}}$, 
        M. Victoria Luzón $^{\text{b}}$,
        Francisco Herrera $^{\text{a,c}}$
}

\begin{document}

\maketitle

\vspace{-1cm}

\begin{centering}
$^\textbf{a}$ \textit{Department of Computer Science and Artificial Intelligence, Andalusian Research Institute in Data Science and Computational Intelligence (DaSCI), University of Granada, Spain}\\
$^\textbf{b}$ \textit{Department of Software Engineering, Andalusian Research Institute in Data Science and Computational Intelligence (DaSCI), University of Granada, Spain} \\
$^\textbf{c}$ \textit{ADIA Lab, AI Maryah Island, Abu Dhabi, United Arab Emirates} \\

\end{centering}

\blfootnote{* Corresponding Author}
\blfootnote{Email addresses:  
\textbf{\texttt{rbnuria@ugr.es}} (Nuria Rodríguez-Barroso),
\textbf{\texttt{luzon@ugr.es}} (M. Victoria Luzón),
\textbf{\texttt{herrera@decsai.ugr.es}} (Francisco Herrera).}
\vspace{0.5cm}

\begin{abstract}

At the same time that artificial intelligence is becoming popular, concern and the need for regulation is growing, \paco{including among other requirements the data privacy}. In this context, Federated Learning is proposed as a solution to data privacy concerns \rab{derived from different source data scenarios} due to its distributed learning. The defense mechanisms proposed in literature are just focused on defending against \final{adversarial} attacks and the performance, leaving aside other important qualities such as explainability, fairness \final{to poor quality clients}, dynamism \final{in terms of attacks configuration} and generality \final{in terms of being resilient against different kinds of attacks}. \final{In this work, we propose RAB\textsuperscript{2}-DEF, a \textbf{r}esilient \textbf{a}gainst \textbf{b}yzantine and \textbf{b}ackdoor attacks which is \textbf{d}ynamic, \textbf{e}xplainable and \textbf{f}air to poor clients using local linear explanations}. We test the performance of RAB\textsuperscript{2}-DEF in image datasets and \rab{both byzantine and backdoor attacks} considering the state-of-the-art defenses and achieve that RAB\textsuperscript{2}-DEF is a proper defense at the same time that it boosts the other qualities towards trustworthy artificial intelligence.

\end{abstract}



\section{Introduction}


\context{Artificial Intelligence (AI) is rapidly transforming many aspects of our lives, has silently crept in, and is already part of our lives. At the same time that we are still unable to even consider the potential of AI in many societal contexts, there is growing concern about the possible negative impacts of AI. In this context, the concept of trustworthy \rab{AI} \cite{thiebes2021trustworthy, diaz2023connecting} arises, based on the pillars of legality, ethics and robustness. In addition, seven technical requirements are set out: (1) human agency 
and oversight, (2) technical robustness and safety, (3) privacy and data governance, (4) transparency, (5) diversity, non-discrimination and fairness, (6) social and environmental well-being, and (7) accountability.}

\context{In this context of regulation  and concern \paco{about trustworthy requirements including  data privacy led by the GDPR \cite{gdpr1} and governance proposals \cite{un2023ai}}, Federated Learning (FL) emerges \cite{kairouz2021advances}. It is presented as a distributed machine learning paradigm in which the data is never shared with other devices. In this way, data privacy, along with proven technical robustness and safety, is ensured. However, although FL is designed to ensure data privacy and robustness, it is still vulnerable to adversarial attacks against both data \cite{erdol2024low} and model integrity~\cite{rodriguez2023survey}.}

\challenge{\final{Poisoning adversarial attacks pose significant threats in FL scenarios. Substantial efforts have been made in the literature to effectively counter these attacks \cite{colosimo2024dynamic, luo2024ganfat}. This has led to the development of several defense mechanisms aimed primarily at enhancing the performance of the federated model and reducing the impact of these attacks. However, most existing strategies suffer from a series of weaknesses} \cite{lyu2022privacy}:
\begin{itemize}
    \item \rab{These methods are designed to be resilient to just one type of attack, with the federated scheme  remaining vulnerable to the rest of attacks.}
    \item \rab{These methods are designed based on some assumptions about the attack configuration, for example, the number of attacks.}
    \item \rab{Because they are based on performance metrics, they can not differ between clients with skewed data and those with adversarial data.} \final{This results in the filtering of poor quality clients, which can harm the robustness of the global model against new data \cite{jiang2024hdhrfl} and deprive these clients of the global learning model, which is unfair to them.}
    \item \rab{These methods are black-box methods and do not provide any explanation about the selection or filtering out of clients.}
\end{itemize}}

\hypothesis{We hypothesize that it is possible to design a general defense mechanism able to address these weaknesses \paco{in a unique proposal}. It has to be generalizable to different kinds of attacks, agnostic and dynamic to changing attack conditions, and show a fair and explainable filtering out of adversarial clients.}

\proposal{This work poses a step further the defense against adversarial attacks and propose \final{defense which is \textbf{r}esilient \textbf{a}gainst \textbf{b}yzantine and \textbf{b}ackdoor attacks, \textbf{d}ynamic, \textbf{e}xplainable and \textbf{f}air to poor clients (RAB\textsuperscript{2}-DEF)}. We \rab{design this defense mechanism inspired in \cite{rodriguez2022dynamic} based on the use of eXplaniable AI (XAI), in particular Local Linear Explanations (LLEs) \cite{sevillano2022revel}.} As we move the focus from performance to LLEs, the key enhancements of this new defense method are as follows:
\begin{itemize}
    \item \textit{\rab{\textbf{R}esilient \textbf{a}gainst \textbf{b}yzantine and \textbf{b}ackdoor attacks}}. As it is not based on performance, it is resilient to both attacks: those that impair performance \rab{(byzantine attacks)} and those that do not \rab{(backdoor attacks)}.
    \item \textit{\textbf{D}ynamic}. It does not fix the number of clients to be filtered out in each round, but it is decided dynamically.
    \item \textit{\textbf{E}xplainable}. As it employs LLEs, visual explanations can be obtained as to why a particular client has or has not been filtered out. \rab{Note that we focus on the RED XAI \final{approach} \cite{biecekposition}, given that we provide model/validation-oriented explanations instead of human/value-oriented explanations.} \rab{Thus, promoting safety and the model behaviour}.
    \item \textit{\textbf{F}air to poor clients}. For the same reason of not being based on performance, it can distinguish between clients with poor performance (poor clients) and adversarial clients. \rab{Although it may be thought that deleting underperforming clients improves the performance of the global model, these clients may possess valuable divergent information for the global model to be able to generalize better to novel information \cite{fang2022robust}, as well as to improve the personalization of those clients \cite{tan2022towards}.}
\end{itemize}}

\evaluation{To asses the performance and \rab{the above-mentioned} desired qualities of RAB\textsuperscript{2}-DEF, we perform several \final{studies}. In particular, we focus on image classification tasks considering three image datasets: Fed-EMNIST, Fashion MNIST and CIFAR-10. Regarding the attack configuration, we consider both byzantine and backdoor attacks, in order to test that our proposal is a general purpose defense. We consider the state-of-the-art baselines against both kinds of attacks. We set up this attack scenario, not only to show the performance of RAB\textsuperscript{2}-DEF as a valid defense, but also to test \final{the two highlighted} qualities, \final{namely,} explainability and fairness to poor clients. }

\paperdesc{The rest of the paper is organized as follows. Section~\ref{sec:background} introduces the concepts needed to follow the rest of the work, including the formal presentation of FL (see Section~\ref{sec:fl}), an introduction to attacks (see Section~\ref{sec:attacks}), explainability (see Section~\ref{sec:explainable_fl}) and fairness (see Section~\ref{sec:fair_fl}) in FL. We deeply explain the core of RAB\textsuperscript{2}-DEF in Section~\ref{sec:gendef}. We specify the experimental setup in Section~\ref{sec:experimental} including: the evaluation datasets in Section~\ref{sec:datasets}, the baselines in Section~\ref{sec:baselines}, the poisoning attacks employed in Section~\ref{sec:poisoning_attacks} and the evaluation metrics in Section~\ref{sec:metrics}. We discuss the experimental results \paco{according to the performance} obtained in Section~\ref{sec:results} and further analyze the proposal \paco{from the point of view of the explainability \rab{in Section~\ref{sec:explainability}} and fairness} in Section~\ref{sec:fairness}. Finally, conclusions are drown in Section~\ref{sec:conclusions}}.

\section{Background}\label{sec:background}

This section provides the background required to follow the rest of the work. 

\subsection{Federated Learning}\label{sec:fl}

FL represents a distributed machine learning paradigm that aims to build machine learning models without directly exchanging training data among participating entities \final{\cite{kairouz2021advances, luzon2024tutorial}}. It operates within a network of clients or data owners, engaging in two primary phases:

\begin{enumerate}
\item \textit{Model training phase:} In this phase, each client collaborates by sharing information without revealing their raw data, thereby jointly training a machine learning model. This model may be hosted by a single client or distributed across multiple clients.

\item \textit{Inference phase:} Subsequently, the clients work together to apply the jointly trained model to process new data instances.
\end{enumerate}

Both phases can operate synchronously or asynchronously, depending on factors such as data availability and the status of the trained model.

It is crucial to note that while privacy preservation is central to this paradigm, another key aspect involves establishing a fair mechanism for distributing the profits generated from the collaboratively trained model.

\rab{After introducing FL as a general idea, a formal FL scenario can be outlined as follows. We consider a group of clients or data owners, denoted as ${C_1, \dots, C_n}$, each having their own local training data ${D_1, \dots, D_n}$. Every client $C_i$ has a local learning model $L_i$, which is defined by the parameters ${L_1, \dots, L_n}$. The primary goal of FL is to develop a global learning model $G$, leveraging the distributed data across clients through a repeated learning process referred to as a ``round of learning".}

\rab{In each round $t$, every client trains its local model via its corresponding local dataset $D^t_i$, which leads to the modification of the local parameters $L^{t}_i$, resulting in updated parameters $\hat{L}^t_i$. Following this, the global parameters $G^t$ are determined by combining the trained local parameters ${\hat{L}^t_1, \dots, \hat{L}^t_n}$ using a predefined federated aggregation function $\Delta$, and the local models are then updated on the basis of the aggregated parameters:}

\major{\begin{equation} \begin{split} 
G^t = \Delta(\hat{L}^t_1,\hat{L}^t_2, \dots, \hat{L}^t_n) \\ 
L^{t+1}_i \leftarrow G^t, \quad \forall i \in {1, \dots, n}. 
\end{split} \label{eq_fl_aggregation} 
\end{equation}}

\rab{This exchange of updates between clients and the server continues until a predefined stopping criterion is reached. Ultimately, the final state of $G$ encapsulates the knowledge learned by the individual clients.}

\begin{figure}[h!]
    \centering
    \includegraphics[width=\linewidth]{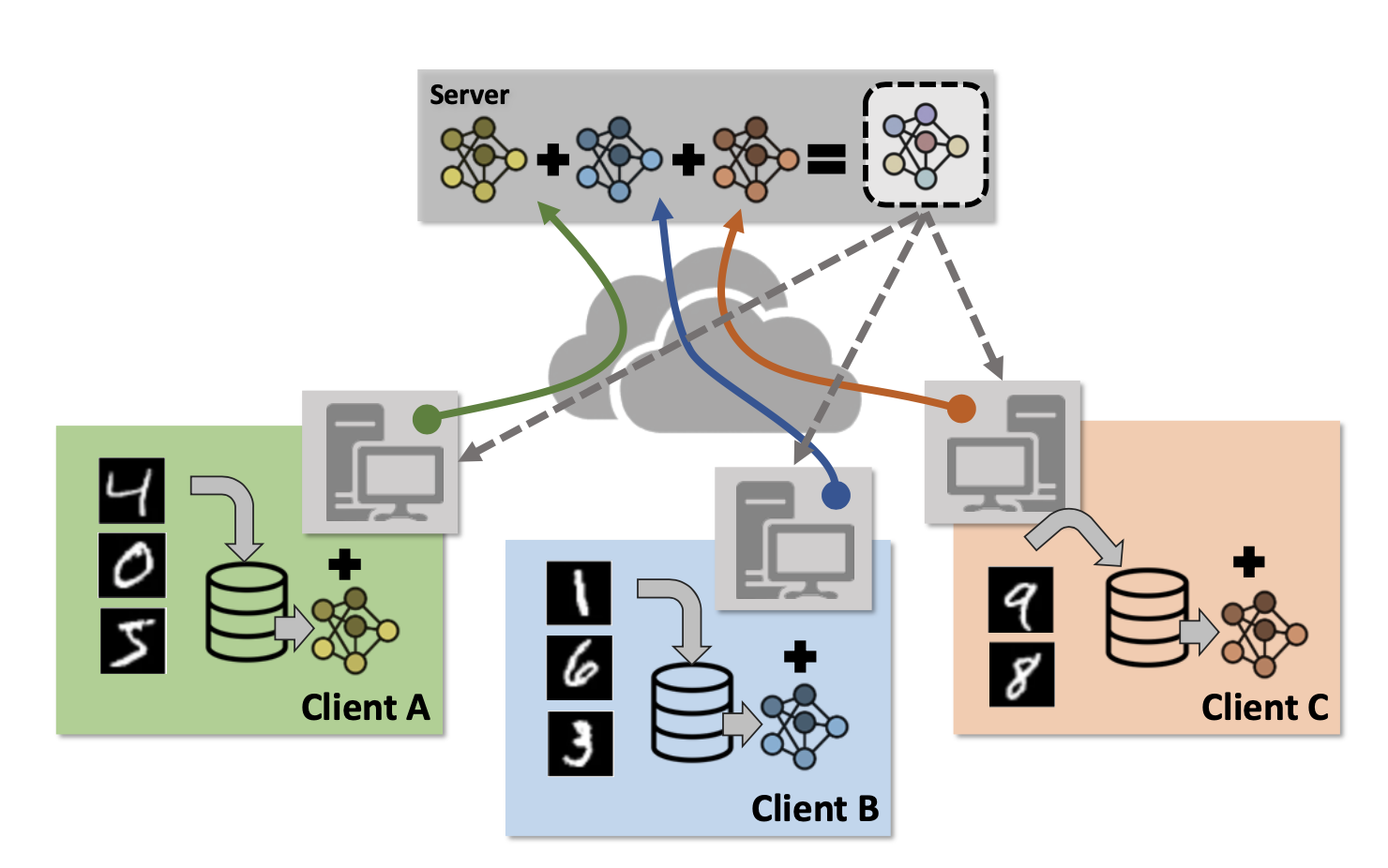}
    \caption{Generic FL scheme, where data is collected at three different clients (A, B and C).}
    \label{fig:fl}
\end{figure}

\final{In Fig. \ref{fig:fl} we present a genericl FL scheme, where model updates are uploaded to a central server  and aggregated to yield a trained global model, which is then delivered downstream to the clients and combined with their local models. As a result, the combined local model leverages knowledge modelled by other client for the same task, while keeping local data private.}

\subsection{Attacks in Federated Learning: Backdoor and Byzantine Threats}\label{sec:attacks}

FL is vulnerable to various adversarial attacks, which can be broadly classified into attacks on the model and privacy attacks \cite{lyu2022privacy, nuriaSurvey2023}. This section focuses on two significant types of model attacks: backdoor attacks and \major{byzantine} attacks.

\rab{\subsubsection{Backdoor attacks in Federated Learning}}

Backdoor attacks \cite{bagdasaryan2020backdoor} involve embedding a hidden, secondary task within the model while preserving its performance on the primary task. These attacks can vary widely based on the specific backdoor task implemented \cite{gong2022backdoor}. One common approach is pattern-key backdoor attacks \cite{wang2020attack}, where attackers introduce a pattern into certain data samples and label these tampered samples with a target label. To enhance the attack's impact and avoid mitigation during aggregation with benign clients' updates, backdoor attacks are often combined with model replacement techniques \cite{bagdasaryan2020backdoor}. This approach amplifies the influence of the adversarial update to ensure it supersedes the benign updates.

Mathematically, let $G^t$ and $L^t_i$ denote the global model and the local model of the $i$-th client at round $t$, respectively, with $n$ clients participating in the round and $\eta$ as the server learning rate. The global model update at round $t$ is given by:
\begin{equation}\label{eq:fedavg}
    G^t = G^{t-1} + \frac{\eta}{n} \sum_{i=1}^n(L_i^t - G^{t-1})
\end{equation}

Assuming a single adversarial client is selected in round $t$, this client attempts to replace the global model $G^t$ with its backdoored model $L^t_{adv}$, optimized for both the primary and backdoor tasks. The adversarial model update is boosted as follows:
\begin{equation}\label{eq:boosting}
    \hat{L}^t_{adv} = \beta(L^t_{adv} - G^{t-1})
\end{equation}
where $\beta = \frac{n}{\eta}$ is the boosting factor. Substituting this into the global model update equation yields:
\begin{equation}\label{eq:replacementexp}
    G^t = G^{t-1} + \frac{\eta}{n} \frac{n}{\eta}(L^t_{adv} - G^{t-1}) + \frac{\eta}{n} \sum_{i=2}^n(L_i^t - G^{t-1})
\end{equation}

Assuming model convergence, $L_i^t - G^{t-1} \approx 0$ for benign clients, thus:
\begin{equation}\label{eq:replacement}
    G^t \approx G^{t-1} + L^t_{adv} - G^{t-1} = L^t_{adv}
\end{equation}

This effectively replaces the global model with the adversarial client's model.

\rab{\subsubsection{Byzantine attacks in Federated Learning}}

Byzantine attacks \cite{shi2022challenges} aim to degrade the model's performance by causing it to behave erratically. These attacks typically involve coordinated actions by adversarial clients to corrupt the learning process through data or model poisoning \cite{tian2022comprehensive}. In data poisoning attacks \cite{yerlikaya2022data}, adversaries introduce harmful patterns into the data, leading to incorrect learning and poisoned models, with label-flipping being a common method \cite{li2022label}. Model poisoning attacks \cite{chen2023tutorial}, on the other hand, involve random modifications to the model's weights, resulting in arbitrary outputs.

Given the potential varying proportion of adversarial clients, these attacks often utilize model replacement techniques \cite{pmlr-v108-bagdasaryan20a} to ensure the adversarial models dominate the global model.

\rab{\subsubsection{Defenses against adversarial attacks in Federated Learning}}

To counteract these threats, numerous defense mechanisms have been proposed \cite{nuriaSurvey2023, shi2022challenges}, which primarily operate on the server due to limited access to the client information. Robust aggregation methods such as MultiKrum \cite{NIPS2017_f4b9ec30}, Bulyan \cite{pmlr-v80-mhamdi18a}, STYX \cite{wen2023styx}, and DDaBA \cite{rodriguez2022dynamic} are designed to filter out malicious updates. However, these defenses can also inadvertently exclude useful information from benign clients with skewed data distributions, thus compromising the principles of fairness and equity essential for trustworthy AI. This can adversely affect the overall performance of the federated model. Recently, advocacy mechanisms have shown promise in maintaining robustness even in highly heterogeneous environments with a significant presence of poor clients \cite{ye2023heterogeneous}.

\subsection{Explainability in Federated Learning}\label{sec:explainable_fl}

The increasing complexity of AI models, particularly in machine learning and deep learning, underscores the necessity of explainability. Explainability, or the ability to understand AI models \rab{according to \cite{arrieta2020explainable}}, is crucial for several reasons. \rab{It can allow} stakeholders to \rab{understand how decisions are made based on explanations}, which is essential for trust and accountability \cite{arrieta2020explainable}. Models that can be easily explained and understood are more likely to be trusted by users, especially in sectors such as healthcare and finance, where decisions can have significant consequences.

Second, explainability aids in the detection and correction of biases within AI systems. Bias in training data can lead to biased outcomes, and without transparency, it is challenging to identify and mitigate these biases. \rab{XAI} enables a better understanding of how models interpret data, making it easier to spot and address potential biases \cite{arrieta2020explainable}. This is essential for developing fair and equitable AI systems that do not perpetuate existing societal inequalities.

However, the distributed nature of FL complicates the process of ensuring explainability. Each client's data may vary significantly, leading to diverse local models that contribute to the global model. This heterogeneity can make it difficult to understand the decision-making process of a global model, as it is influenced by a multitude of local datasets and training processes.

Despite these challenges, integrating explainability into FL is essential. This helps in understanding the contributions of individual client models to the global model, ensuring that the aggregated model is robust and free from biases present in any single client's data. Moreover, explainability in FL can foster trust among participants, as they can gain insights into how their data are being used and how it influences the global model \cite{barcena2022fed, biecekposition}.

\subsection{Fairness in Federated Learning}\label{sec:fair_fl}

Fairness and FL are critical components in the advancement of responsible AI. Fairness ensures that \rab{AI assisted} decision-making systems do not perpetuate historical biases or discriminate against minority groups, thus promoting the ethical and responsible use of technology \cite{lyu2020collaborative}. In the context of FL, significant research efforts are dedicated to addressing fairness concerns. This includes developing algorithms that ensure equitable performance across diverse data sources and demographic groups, as well as techniques to identify and mitigate bias during the federated training process \cite{yu2020fairness}. Researchers are also exploring methods to measure and improve fairness in federated settings, such as fairness-aware aggregation techniques and bias correction mechanisms \cite{ezzeldin2023fairfed}. By prioritizing fairness in FL, we can ensure that these distributed models not only protect user privacy but also deliver equitable results for all clients \rab{considering that the clients can be affected by unfair decisions.}.

In this paper, we use the concept of the \textit{poor client} as a \textit{client} who has a skewed distribution of data. This skewed distribution can be in terms of features or in terms of labels. Throughout this paper, we will refer to \textit{fairness} in terms of participation in the model. In many situations, adversarial defense mechanisms filter out clients based on their performance, even filtering out poor clients as well, which is unfair. 

\rab{\section{RAB\textsuperscript{2}-DEF: Dynamic, explainable, and fair defense for poor clients against byzantine and backdoor attacks}\label{sec:gendef}}

As stated in the Introduction, the main motivation is to develop a defense against adversarial attacks, that is,
\begin{itemize}
    \item \textit{\rab{\textbf{R}esilient \textbf{a}gainst \textbf{b}yzantine and \textbf{b}ackdoor attacks}}: \rab{It does not depend on the performance of the clients.}
    \item \textit{\textbf{D}ynamic}: It is able to adapt to different number of adversarial clients.
    \item \textit{\textbf{E}xplainable}: It can explain why a client has been discarded or not.
    \item \textit{\textbf{F}air to poor clients}: It can differ between poor clients (with skewed data distributions) and adversarial clients, not discarding the poor ones.
\end{itemize}

To create a defense strategy that meets the criteria of generality, dynamism, explainability, and fairness, \rab{we design RAB\textsuperscript{2}-DEF, a dynamic and explainable defense against byzantine and backdoor attacks fair to poor clients} \major{based on the improvement of the previous proposal DDaBA \cite{rodriguez2022dynamic}}. \rab{For that purpose, we set a small test set located at the central server to classify clients as either \emph{adversarial} or \emph{non-adversarial} based on XAI techniques. \paco{It} includes the following components:}
\begin{enumerate}
    \item \textit{LLEs-based induced ordering function for client model updates}: This function ranks clients based on the \rab{LLEs} over the server's test data. \rab{Our hypothesis is that this ordering not only sustains a good robustness against attacks, but also endows the server with the ability to explain why a certain client is identified as adversarial and hence filtered out from the aggregation. For that purpose, we employ the LLEs to measure how different the update of a specific client is from the rest of the clients.}
    \item \textit{Dynamic linguistic quantifier for weighting the contribution of clients}: This function assigns weights to each client's contribution, giving a weight of zero to those deemed adversarial, while distributing the remaining weights such that the top-performing clients have twice the contribution of the others. \rab{For that purpose, we define a step-wise function based on data distribution of the clients model updates sorted using the LLEs-based induced ordering function}.
    \item \textit{Defense based on federated aggregation}: The defense employs a weighted aggregation operator, with each client's contribution determined by the dynamic linguistic quantifier.
\end{enumerate}



\paragraph{\rab{LLEs-based induced ordering function}}
\rab{Formally}, for each client we define a \textit{LLEs ordering function} for each local update $L_i$ as follows:
\begin{equation}\label{eq:fLE}
    f_{LE}(L_i) = \sum_{\mathbf{x}_v \in \mathbf{X}_v} S_C(\mathbf{A}^p_{i,v}, \mathbf{A}^p_{j,v}), \quad \forall L_j \in \mathcal{L},
\end{equation}
where $\mathcal{L} = \{L_1, \dots, L_n\}$ denotes all the model updates for the $n$ clients in the federation; $S_C(\cdot,\cdot)$ denotes average cosine similarity; $\mathbf{X}_v$ is the validation dataset allocated in the server, and $\mathbf{A}^p_{i,v}$ the importance matrix over the probability spaces of $L_i$ computed for validation instance $\mathbf{x}_v$. Under the assumption that the local updates will converge to a common solution, we define a random variable $X_i^{f_{LE}}$ as:
\begin{equation}
X_i^{f_{LE}} = \max_{i\in\{1,\ldots,n\}}\{f_{LE}(L_i)\} - f_{LE}(L_i),
\end{equation}
which will approximate an exponential distribution with rate $\lambda$. 

\paragraph{\rab{Dynamic linguistic quantifier}} We define the \rab{dynamic linguistic quantifier} weighting $w_i^{(a,b,c,y_b)}$ assigned to each model update $L_i$ \rab{as a step-wise function} depending on parameters $a$, $b$, $c$ and $y_b$ as follows:
\begin{equation}
    w_i^{(a,b,c,y_b)} = Q_{a,b,c,y_b} \left( \frac{i}{n} \right) - Q_{a,bc,y_b} \left( \frac{i-1}{n} \right)
    \label{eq_q_weight_calculation_1_paper}
\end{equation}
where $a,b,c \in \mathbb{R}[0,1]$ satisfying $0 \leq a \leq b \leq c \leq 1$, and:
\begin{itemize}
    \item $a=0$.
    \item $b$ is the proportion of clients that verify:
    \begin{equation}
    X_i^{f_{LE}} \leq \frac{\ln(10/9)}{\lambda},
    \end{equation}
    where $\lambda = 1/E[X_i^{f_{LE}}]$ (inverse expected value of $X_i^{f_{LE}}$).

    \item $c = 1-\hat{c}$, with $\hat{c}$ being the proportion of clients verifying:
    \begin{equation}
        X_i^{f_{LE}} \geq Q_3 + 1.5\cdot IQR = \frac{\ln(4)}{\lambda} + 1.5 \frac{\ln(3)}{\lambda},
    \end{equation}
    with $Q_3$ and $IQR$ denoting third quartile and interquartile range, respectively.

    \item $y_b = 2|Top|/(2|Top|-|Rest|)$ where $|Top| = b \cdot n$ and $|Rest| = (c-b)\cdot n$.

    \item $Q_{a,b,c,y_b}(x)$ is the \rab{step-wise function} defined as:
    \begin{equation}\label{eq_dynamic_quantifier}
    Q_{a,b,c,y_b}(x) = 
    \begin{dcases}
        0 & 0\leq x \leq a\\
        \frac{x-a}{b-a} \cdot  y_b & a \leq x \leq b \\
        \frac{x-b}{c-b} \cdot (1-y_b) + y_b & b \leq x \leq c \\
        1 & c \leq x \leq 1
    \end{dcases}
\end{equation}
\end{itemize}

\paragraph{\rab{Defense based on federated aggregation}}
We define the proposed RAB\textsuperscript{2}-DEF defense strategy based on the following aggregation operator:
\begin{equation} \label{eq:GEN-DEFddaba-aggoperator}
     \text{RAB\textsuperscript{2}-DEF}(\{L^t_1, L^t_2, \ldots, L^t_n\}, \mathbf{X}_v) = \sum_{i=1}^n w_i^{(a,b,c,y_b)} L^t_i,
\end{equation}
where $w_i^{(a,b,c,y_b)}$ is defined in Expression \eqref{eq_q_weight_calculation_1_paper}, and $L_i^t$ the local model update of the client $i\in\{1, \ldots, n\}$. 

\paragraph{\major{Conceptual differences with DDaBA}}

\major{The proposed method, RAB\textsuperscript{2}-DEF, is an improvement on the previous proposal DDaBA for byzantine attacks, also based on IOWA operators but using a client ordering function based on client performance in terms of accuracy. The main difference lies in the ordering function, which in this case is based on LLEs. This is an adaptation that has a great impact, the conceptual differences are important to note}:

\major{\begin{itemize}
    \item \textbf{Scope of the defense:} Backdoor attacks, unlike byzantine attacks, do not result in a detriment to the performance of the model in the original task. For this reason, performance-based defenses are not resilient to backdoor attacks. However,  RAB\textsuperscript{2}-DEF, being based on LLEs, will be able to identify these backdoor adversarial clients. 
    \item \textbf{Fairness:} Performance-based defenses are unable to differentiate between adversarial clients and poor clients (with skewed data distributions) since both produce a loss of performance. We claim that  RAB\textsuperscript{2}-DEF, being based on LLEs, will be able to differentiate this type of clients, allowing poor clients to participate in the aggregation, making the decision fairer and the model more robust.
    \item \textbf{Explainability:} Performance-based defenses simply discard clients based on their performance, without being able to give any further explanation. In contrast, RAB\textsuperscript{2}-DEF provides visual explanations as to why a client has been discarded from aggregation or not.
\end{itemize}}

\section{Experimental setup} \label{sec:experimental}

In this section we detail the experimental setup employed to test our proposal. In the following, we detail the evaluation datasets (see Section~\ref{sec:datasets}), baselines (see Section~\ref{sec:baselines}) and poisoning attacks (see Section~\ref{sec:poisoning_attacks}).

\subsection{Evaluation datasets}\label{sec:datasets}

Since attacks and defenses are independent of the classification task, we can focus on image classification problems, which are the most common in studies of poisoning attacks, without losing generality. The considered datasets are as follows:
\begin{itemize}
    \item The Fed-EMNIST dataset \cite{lecun98}. The EMNIST Digits contains a balanced subset of the DIGITS dataset containing 28,000 samples of each digit. The dataset consists of 280,000 samples, in which 240,000 are training samples and 40,000 test samples. We use its federated version by identifying each client with an original writer.

    \item The Fashion MNIST \cite{DBLP:journals/corr/abs-1708-07747}, which contains a balanced subset of 10 different classes containing 7,000 samples of each class. Hence, the dataset consists of 70,000 samples, which 60,000 are training samples and 10,000 test samples. We fix the number of clients to 500.
    
    \item The CIFAR-10 dataset is a labeled subset of the 80 million tiny image dataset \cite{4531741}. It consists of 60,000 32$\times$32 color images in 10 classes, with 6,000 images per class. There are 50,000 training images and 10,000 test images, which correspond to 1,000 images of each class. We set the number of clients to 100.
\end{itemize}

Due to the fact that we need some data in the server to apply the defense strategy  -- \textit{validation dataset} $\textbf{X}_{v}$ defined in \eqref{eq:fLE} -- we employ 20\% of the test set for this. This yields the evaluation datasets shown in Table \ref{tab:4_nonad_emnist}.

\begin{table}[!h]
\centering
\scriptsize
\caption{Sizes of the training, validation and test partitions of Fed-EMNIST, Fashion MNIST and CIFAR-10 datasets.}
\label{tab:datasets}
\begin{tabular}{lccc}
\toprule
 & \textbf{Training} & \textbf{Validation} ($\mathbf{X}_v$) & \textbf{Test}\\
\midrule
\textbf{Fed-EMNIST} & 240,000 & 8,000 & 32,000\\
\textbf{Fashion MNIST} & 60,000 & 2,000 & 8,000\\
\textbf{CIFAR-10} & 60,000 & 2,000 & 8,000\\
\bottomrule
\end{tabular}
\label{tab:4_nonad_emnist}
\end{table}

\subsection{Baselines}\label{sec:baselines}

To test the resilience against adversarial attacks of our proposal, we use the following baselines:

\paragraph{Common baselines} We employ two simple baselines to represent the starting point.

\begin{itemize}
    \item \textit{Median} \cite{chen2017distributed}: the average is changed with the median in the aggregation process, which is more robust with respect to the extreme values.
    \item \textit{Trimmed-mean} \cite{DBLP:journals/corr/abs-1803-01498}, which uses a more robust version of the mean that consists of eliminating a fixed percentage (15\%) of extreme values, both above and below the data distribution.
\end{itemize}

\paragraph{Baselines to byzantine attacks} \rab{We employ state-of-the-art baseliens against byzantine attacks, such as:}

\begin{itemize}
\item \textit{Multikrum} \cite{NIPS2017_f4b9ec30} sorts the clients according to the geometric distances of their local model updates. After that, it employs an aggregation parameter, which specifies the number of clients (20) to participate in the aggregation process (the best ones after being sorted). 
\item \textit{Bulyan} \cite{pmlr-v80-mhamdi18a} combines Multikrum and the trimmed-mean. That is, it sorts the clients according to their geometric distances, and filters out a fraction (15\%) of the clients falling in the tails of the sorted distribution of clients. After that, it computes the aggregation of the remaining clients.
\end{itemize}

\paragraph{Baselines to backdoor attacks} We now specify the specific baselines against backdoor attacks. These baselines take into account the double goal of backdoor attacks in order to defend against them.

\begin{itemize}

    \item \textit{Norm Clipping of updates} \cite{DBLP:journals/corr/abs-1911-07963}. Since the boosting factor produce large norms in backdoor attacks model updates, norm clipping of updates is commonly used as a defense mechanism against these attacks. It involves clipping the update by dividing it by the appropriate scalar if it exceeds a fixed threshold $M$, as in Equation \ref{normclipping}, where $\Delta L_i^{t} = L_i^{t+1} - G^t$.
    
    \begin{equation}\label{normclipping}
        G^{t+1} = G^t + \frac{\eta}{n}\sum_{i=1}^n \frac{\Delta L_i^{t}}{\max(1, ||\Delta L_i^{t}||_2/\mathbf{M} )}
    \end{equation}
    
    \item \textit{Weak Differential Privacy (WDP)} \cite{DBLP:journals/corr/abs-1911-07963}. This defense is based on Differential Privacy \cite{dwork2006differential}, widely used to prevent against backdoor attacks \cite{bagdasaryan2020backdoor}. This mechanism involves applying norm techniques combined with a small amount of Gaussian noise as a function of $\sigma$ according to Equation \ref{wdp}.
    
    \begin{equation}\label{wdp}
        G^{t+1} = G^t + \frac{\eta}{n}\sum_{i=1}^n \frac{\Delta L_i^{t}}{\max(1, ||\Delta L_i^{t}||_2/\mathbf{M} )} + \mathcal{N}(0, \frac{\mathbf{\sigma M}}{n})
    \end{equation}
    
    \item \textit{Robust Learning Rate (RLR)} \cite{ozdayi2020defending}. They determine the direction of the update for each dimension using the signs of the updates and a threshold parameter $\theta$. If the sum of the signs of the updates is less than a fixed $\theta$, they change the direction of the update by multiplying it by $-1$. They assert that this defense can be combined with the two previous ones by applying norm clipping and noise addition to the modified models' updates, resulting in better performance.
\end{itemize}

\subsection{Poisoning attacks}\label{sec:poisoning_attacks}

In the following, we specify the poisoning attacks implemented for the experimental results. We employ both data and model poisoning attacks, and both byzantine and backdoor attacks.

\paragraph{Byzantine attacks} These attacks consist of randomly poisoning some part of the data or model updates. In particular, we implement:
\begin{itemize}
    \item \textit{Label-flipping attack} \cite{DBLP:journals/corr/abs-2007-08432}, which involves randomly altering the labels of the adversarial clients. Consequently, these clients learn from poisoned data, which they then transmit to the server for aggregation, thereby compromising the aggregated model.

    \item \textit{Random weights} \cite{DBLP:journals/corr/abs-1911-12560}, which is a model poisoning attack consisting of randomly producing the model updates assigned to each adversarial client. 
\end{itemize}

\paragraph{Backdoor attacks} These attacks involve of injecting a secondary task. For this purpose, we implement pattern-key attacks, which are based on identifying the samples poisoned with some pattern with the target label. For the sake of showing that the performance of the defense is agnostic of the pattern-key, we employ different patterns: 
\begin{itemize}
    \item A black cross of length 3 for Fed-EMNIST and Fashion MNIST.
    \item A 5x5 white square for CIFAR-10.
\end{itemize}

\begin{figure}[h!]
\centering
\begin{subfigure}{.25\linewidth}
  \centering
  \includegraphics[width=0.9\linewidth]{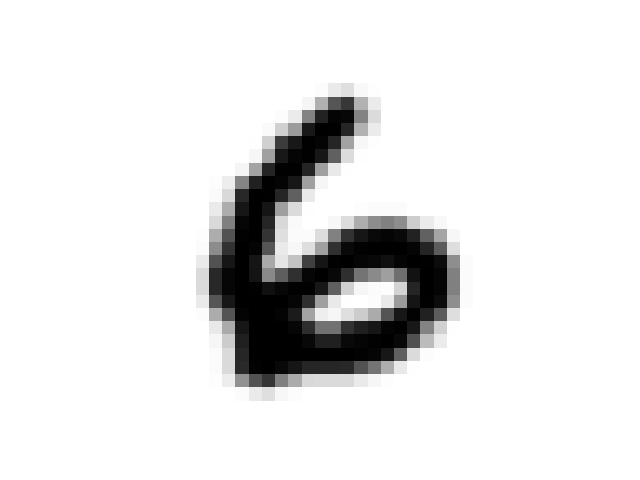}
  \caption{Original image.}
\end{subfigure}
\begin{subfigure}{.25\linewidth}
  \centering
  \includegraphics[width=0.9\linewidth]{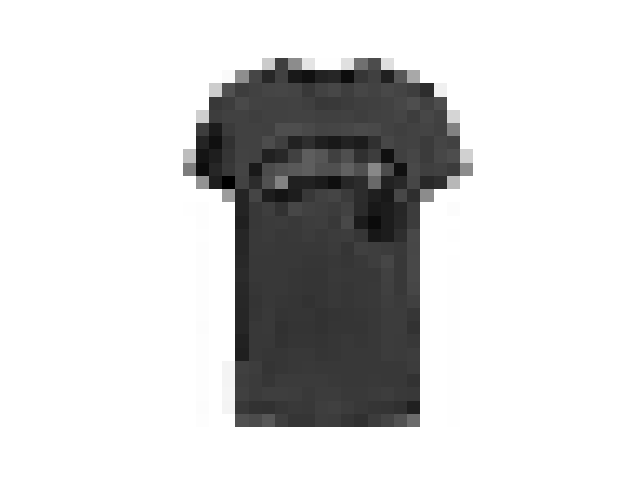}
  \caption{Original image.}
\end{subfigure}
\begin{subfigure}{.25\linewidth}
  \centering
  \includegraphics[width=0.9\linewidth]{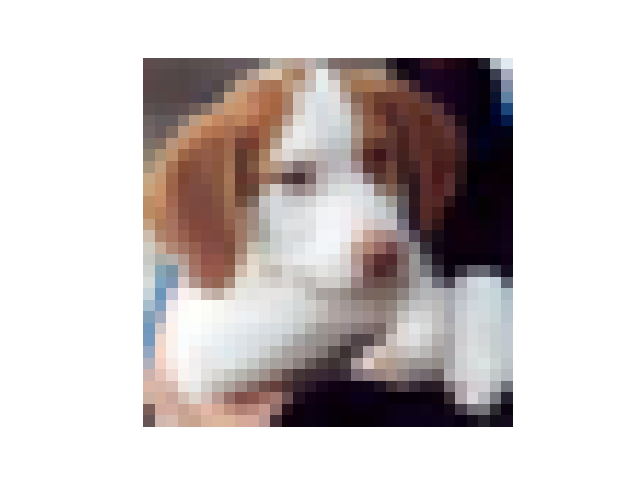}
  \caption{Original image.}
\end{subfigure}
\vskip\baselineskip
\begin{subfigure}{.25\linewidth}
  \centering
  \includegraphics[width=0.9\linewidth]{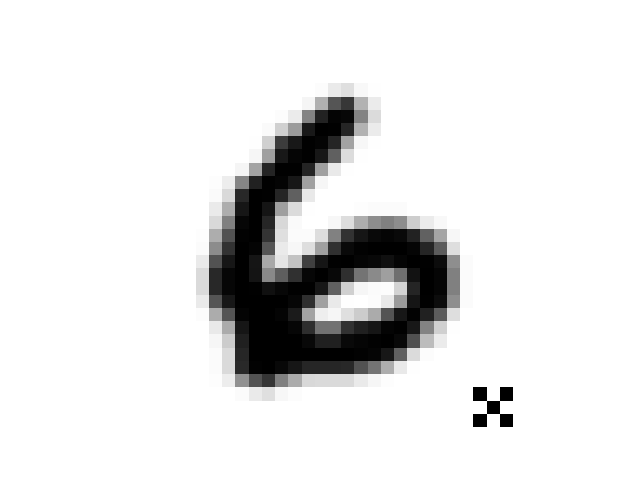}
  \caption{Poisoned image.}
\end{subfigure}
\begin{subfigure}{.25\linewidth}
  \centering
  \includegraphics[width=0.9\linewidth]{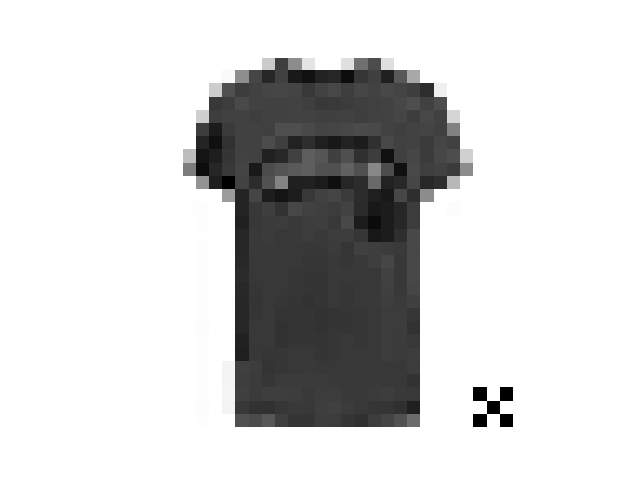}
  \caption{Poisoned image.}
\end{subfigure}
\begin{subfigure}{.25\linewidth}
  \centering
  \includegraphics[width=0.9\linewidth]{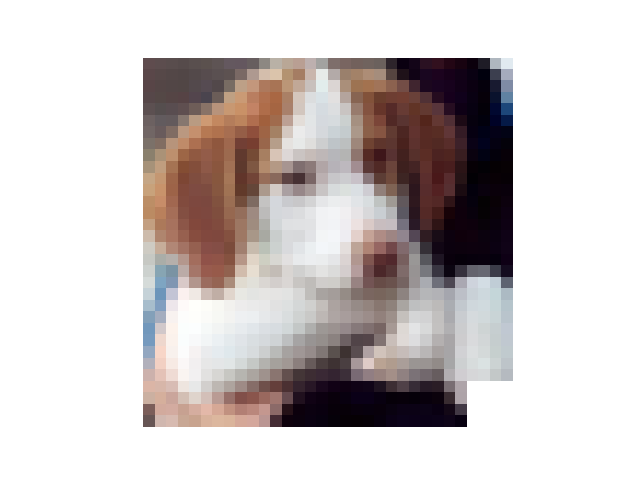}
  \caption{Poisoned image.}
\end{subfigure}
\vskip\baselineskip
\caption{Examples of original (a, b and c) and backdoored (d, e and f) samples.}
\label{fig:patterns}
\end{figure}

\rab{In Fig. \ref{fig:patterns} we show the selected patterns for the backdoor attacks. From left to right: (1) a black cross pattern of length 3 in the bottom-right corner for a Fed-EMNIST instance; a black cross pattern of length 3 for a Fashion MNIST instance; and (3) a white squared pattern in the bottom-right corner for a CIFAR instance.}

\subsection{Evaluation metrics}\label{sec:metrics}

As the objectives of byzantine and backdoor attacks are different, we measure the performance of each kind of defense in different ways.

\paragraph{Evaluation metrics for byzantine attacks} As the goal of byzantine attacks is to impair the performance of the global model, we employ the average test accuracy ($accuracy$) of the global model. The higher it is, the better the defense, as the more it is mitigating the effect of the attack.

\paragraph{Evaluation metrics for backdoor attacks} As backdoor attacks have a double goal (to inject the secondary task while maintaining the performance in the original one) we use both metrics: (1) \textit{Original task test ($Original$)}, corresponding with the test accuracy in the original task; and (2) \textit{Backdoor task test ($Backdoor$)}, corresponding with the test accuracy in the backdoor task. Clearly, the best defense is the one which achieves the highest \textit{original} accuracy and the lowest \textit{backdoor} accuracy.

\paco{In addition, for the purpose of a more comprehensive analysis of the proposal, \rab{we use the following metrics:}}

\paco{\paragraph{Evaluation metrics to explainability} The explainability of the proposal is going to be measured using visual explanations according to the importance of each pixel.. For that reason, in Section \ref{sec:explainability} we depict images where the importance of each pixel is measured in a range of greys where total white represents maximum importance and black represents minimum importance.}

\paco{\paragraph{Evaluation metrics to fairness} In order to measure the fairness of the proposal in Section \ref{sec:fairness}, we count the minimum (Min), maximum (Max) and average (Avg) number of both adversarial and poor clients discarded along the rounds of learning. Fairness will be substantiated when fewer poor clients are discarded.}

\section{Experimental results}\label{sec:results}

In this section we discuss the experimental results obtained by RAB\textsuperscript{2}-DEF in comparison with the baselines in the adversarial attacks specified. \paco{Currently, we focus only on the precision in terms of accuracy, in the following we perform further analysis.} We show the results against byzantine attacks in Section \ref{sec:byzantine} and the results against backdoor attacks in Section \ref{sec:backdoor}. \rab{In the first row we also show the average accuracy of \textit{FedAvg} without any attack. The best result for each of the scenarios is highlighted in bold.}

\subsection{Results against byzantine attacks}\label{sec:byzantine}

In the following we report the results obtained by RAB\textsuperscript{2}-DEF and all the baselines considered un both the \textit{label-flipping} (see Table \ref{tab:label-flipping}) and \textit{random weights} (see Table \ref{tab:random-weights}) byzantine adversarial attacks. 

\begin{table}[ht!]
\caption{Mean accuracy results for the \textit{label-flipping} byzantine attack.}
\label{tab:label-flipping}
\begin{center}
\small
\resizebox{.85 \linewidth}{!}{
\begin{tabular}{lccc}
\toprule
 & \textbf{Fed-EMNIST} & \textbf{Fashion MNIST} & \textbf{CIFAR-10}\\
 \toprule
\textbf{No attack} & 0,9657 & 0,8719 & 0,8357\\
\midrule
\textbf{FedAvg} & 0,4210 & 0,3661 & 0,1436\\
\midrule
\textbf{Median} & 0,9096 & 0,8396 & 0,8087\\
\textbf{Trimmed Mean} & 0,8841 & 0,8471 & 0,7552\\
\midrule
\textbf{MultiKrum} & 0,9270 & 0,8433 & 0,8467\\
\textbf{Bulyan} & 0,9423 & 0,8665 & 0,8475\\
\textbf{DDaBA} & 0,9853 & 0,8832 & 0,8503\\
\midrule
\textbf{RAB\textsuperscript{2}-DEF} & \textbf{0,9855} & \textbf{0,8835} & \textbf{0,8510}\\ 
\bottomrule
\end{tabular}}
\end{center}
\end{table}

\begin{table}[ht!]
\caption{Mean accuracy results for the \textit{random weights} byzantine attack.}
\label{tab:random-weights}
\begin{center}
\small
\resizebox{.85 \linewidth}{!}{
\begin{tabular}{lccc}
\toprule
  & \textbf{Fed-EMNIST} & \textbf{Fashion MNIST} & \textbf{CIFAR-10}\\
 \toprule
\textbf{No attack} & 0,9657 & 0,8719 & 0,8357\\
\midrule
\textbf{FedAvg} & 0,0994 & 0,1016 & 0,0994\\
\midrule
\textbf{Median} & 0,9295 & 0,8620 & 0,8557\\
\textbf{Trimmed Mean} & 0,1052 & 0,1021 & 0,0994\\
\midrule
\textbf{MultiKrum} & 0,9564 & 0,8661 & 0,8393\\
\textbf{Bulyan} & 0,9399 & 0,8678 & 0,8413\\
\textbf{DDaBA} & 0,9650 & 0,8734 & \textbf{0,8634}\\
\midrule
\textbf{RAB\textsuperscript{2}-DEF} & \textbf{0,9671} & \textbf{0,8752} & 0,8603\\ 
\bottomrule
\end{tabular}}
\end{center}
\end{table}

Although each table shows the experimental results for a different kind of attack, the same conclusions are obtained from each of them:

\begin{itemize}
\item \major{Compared with the ``No attack'' scenario, RAB\textsuperscript{2}-DEF performs better. This is due to the weighted aggregation that is performed by giving more weight to the clients considered “Top” (see equations \ref{eq_q_weight_calculation_1_paper}, \ref{eq_dynamic_quantifier} and \ref{eq:GEN-DEFddaba-aggoperator}). Although RAB\textsuperscript{2}-DEF does not filter out poor or skewed clients like DDaBA, it does give higher weights to clients it considers the top ones. This may causes an improvement in global performance compared to FedAvg in a scenario without attacks, which aggregates all clients following an unweighted average.}
\item Compared to baselines, RAB\textsuperscript{2}-DEF produces good results, outperforming them in both attacks.
\item Compared to DDaBA, we find that RAB\textsuperscript{2}-DEF consistently provides competitive results even when RAB\textsuperscript{2}-DEF is not designed to optimize the performance. In some scenarios DDaBA performs slightly better yet not significantly superior to RAB\textsuperscript{2}-DEF.
\end{itemize}

These \rab{accuracy-based} findings validate that RAB\textsuperscript{2}-DEF is a robust defense against byzantine poisoning attacks. Although its performance margins are slightly close to other counterparts in the benchmark, RAB\textsuperscript{2}-DEF \major{is designed to be resilient to other kinds of attacks as we see in the following section.}

\subsection{Results against backdoor attacks}\label{sec:backdoor}

In the following, we test wether RAB\textsuperscript{2}-DEF is a valid defense against backdoor attacks \major{based on the assumption that it is not a performance-based defense.} We present the results in Table \ref{tab:backdoor-attacks}. 

\begin{table*}[ht!]
\caption{Mean accuracy results for the \textit{pattern-key} backdoor attack. }
\label{tab:backdoor-attacks}
\begin{center}
\resizebox{.85 \linewidth}{!}{\begin{tabular}{lrrrrrr}
\toprule
  & \multicolumn{2}{c}{\textbf{Fed-EMNIST}}  & \multicolumn{2}{c}{\textbf{Fashion MNIST}}  &  \multicolumn{2}{c}{\textbf{CIFAR-10}}\\
  & \textit{Original} & \textit{Backdoor} & \textit{Original} & \textit{Backdoor} & \textit{Original} & \textit{Backdoor} \\
 \toprule
\textbf{No attack} & 0,9657 & - & 0,8719 & - & 0,8357 & -\\
\midrule
\textbf{FedAvg} & 0,9598 & 1.0  & 0,8671 & 0,99 & 0,8329 &  0,99\\
\textbf{DDaBA} & 0,9603 & 0,2739 & 0,8599 & 0,3135 & 0,8352 & 0,2893\\
\midrule
\textbf{Median} & 0,9235 & 0,0158 & 0,8378 & 0,0203 & 0,8197 & 0,0174\\
\textbf{Trimmed Mean} & 0,9301 & 0,0203 & 0,8653 & 0,0193 & 0,8271 & 0,0186\\
\midrule
\textbf{NormClip} & 0,9587 & 0,0553 & 0,8561 & 0,0712 & 0,8291 & 0,0801\\
\textbf{WDP} & 0,9357 & 0,0921 & 0,8653 & 0,0698 & 0,8332 & 0,0793\\
\textbf{RLR} & 0,9265 & \textbf{0,0089} &  0,8599 & 0,0091 & 0,8239 & 0,0095\\
\midrule
\textbf{RAB\textsuperscript{2}-DEF} & \textbf{0,9612} & 0,0101 & \textbf{0,8693} & \textbf{0,0088} & \textbf{0,8597} & \textbf{0,0093}\\ 
\bottomrule
\end{tabular}}
\end{center}
\end{table*}

\begin{itemize}
\item Compared to baselines, RAB\textsuperscript{2}-DEF produces good results, outperforming all considered baselines \major{and confirming that it is also resilient against backdoor attacks}.

\item \major{Regarding its comparison to DDaBA, we confirm the hypothesis that switching the ordering function from a performance-based to an LLE-based ordering function expands the scope of the defense. This means that RAB\textsuperscript{2}-DEF is able to defend against backdoor attacks (according to \textit{Backdoor} columns) without any performance penalty compared to performance-based metrics such as DDaBA (according to \textit{Original} columns).}
\end{itemize}

\major{These accuracy-based findings in the backdoor attack scenario validate that RAB\textsuperscript{2}-DEF is a robust defense against backdoor attacks. This demonstrates the first difference from the previous DDaBA proposal, which was the broadening of the scope of attacks to which it is resilient. The other two improvements, which are the enhancements in fairness and explainaiblity, will be examined in the following sections.}


\rab{\section{Analysis on explainability}\label{sec:explainability}}

\begin{figure}[h!]
\centering
\begin{subfigure}{.25\linewidth}
    \centering
    \includegraphics[width=\linewidth]{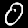}
    \caption{Validation image.}
    \label{sfig:initial}
\end{subfigure}
\begin{subfigure}{.25\linewidth}
    \centering
    \includegraphics[width=\linewidth]{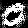}
    \caption{Regular client.}
    \label{sfig:initial}
\end{subfigure}
    \begin{subfigure}{.25\linewidth}
    \centering
    \includegraphics[width=\linewidth]{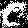}
    \caption{Poor client.}
    \label{sfig:initial}
\end{subfigure}
    \begin{subfigure}{.25\linewidth}
    \centering
    \includegraphics[width=\linewidth]{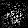}
    \caption{Random weights.}
    \label{sfig:initial}
\end{subfigure}
\begin{subfigure}{.25\linewidth}
    \centering
    \includegraphics[width=\linewidth]{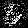}
    \caption{Label-flipping.}
    \label{sfig:initial}
\end{subfigure}
\begin{subfigure}{.25\linewidth}
    \centering
    \includegraphics[width=\linewidth]{images/client_0_image_0.png}
    \caption{Backdoor attack.}
    \label{sfig:initial}
\end{subfigure}
\caption{Example of an original image (a), and the explanations in terms of feature importance of (b) a regular client; (c) a poor client; (d) an adversarial client implementing a random weights attack; (e) an adversarial client implementing a label-flipping attack; and (f) an adversarial client implementing a cross-pattern backdoor attack.}
\label{fig:instance}
\end{figure}

The analysis of explainability delves into a fundamental characteristic of RAB\textsuperscript{2}-DEF: the client selection process based on LLEs inherently provides an explanation for why a client is either included or excluded. Since LLEs rely on feature importance, we can illustrate the significance of each feature within an image and evaluate if the model is concentrating on image areas that intuitively correspond with the predicted categories. As explained in Section \ref{sec:gendef}, RAB\textsuperscript{2}-DEF utilizes the resemblance among these explanations to determine whether to retain or exclude a client's model from the aggregation process. Consequently, visually examining the explanations linked to various client models for a validation sample (or a collection of samples) can assist a client in comprehending why its model is either included in or excluded from the aggregation, thus revealing the aggregation criteria embedded within the proposed defense strategy.

Examples of LLEs for the different attacks under consideration and a validation image of the MNIST dataset (digit $0$) are shown in Figs. \ref{fig:instance}.a to \ref{fig:instance}.e. Although the corresponding explanation for the regular client (Fig. \ref{fig:instance}.b) highlights relevant zones for the image label (the contour of the digit $0$) more clearly than for the poor client (Fig. \ref{fig:instance}.c), the explanations of both clients match fairly closely the informative regions of this particular digit. In contrast, the visual cues that the adversarial client model (Figs. \ref{fig:instance}.d and \ref{fig:instance}.e) considers important are scattered randomly over the image. This would lead to a high dissimilarity of such feature importance map w.r.t. those of the rest of clients in the aggregation, and would ultimately yield their models being filtered out. Finally, if we compare between the three adversarial attacks considered, we find that in the data poisoning attack (\textit{label-flipping}), explanations become slightly more noticeable in some parts of the contour of digit $0$, while in the model poisoning attack (\textit{random weights}) the explanation fails to match any of the shape particularities of the digit. Nevertheless, differences in the three cases with the poor and normal clients are large enough for them to be distinguishable from the adversarial clients in all situations.

\rab{\section{Analysis on fairness}\label{sec:fairness}}

We begin our analysis by evaluating whether RAB\textsuperscript{2}-DEF ensures fairness for all clients. \rab{Other accuracy-based baselines} lacks fairness as its filtering criterion may exclude clients with a poor (skewed) distribution of data. This unfair exclusion can negatively impact both these disadvantaged clients and the global model, as such clients may hold relevant information for other clients. \rab{We perform two analyses: (1) we count the number of adversarial and poor clients discarded, and (2) we analyse the performance of the poor clients in the original task.}

\subsection{Comparison in terms of adversarial and poor clients discarded}

To assess this, we count the number of adversarial and poor clients discarded in each learning round and report the minimum, maximum, and average number of discarded adversarial and poor clients in Tables \ref{tab:adv} and \ref{tab:poor}. \rab{As the aim is to demonstrate that RAB\textsuperscript{2}-DEF is an improvement in terms of fairness over DDaBA (based on accuracy), we only consider these two methods in the analysis.}

\begin{table*}[!h]
\centering
\caption{Minimum (\textbf{Min}), maximum (\textbf{Max}) and average (\textbf{Avg}) number of adversarial clients discarded by DDaBA and RAB\textsuperscript{2}-DEF throughout the learning rounds for the \textit{label-flipping} and \textit{random weights} attacks.}
\resizebox{\linewidth}{!}{\begin{tabular}{llccccccccccccc}
\toprule
 & & \multicolumn{3}{c}{\textbf{Fed-EMNIST}} & & \multicolumn{3}{c}{\textbf{Fashion MNIST}}  & & \multicolumn{3}{c}{\textbf{CIFAR-10}}\\
 \cmidrule{3-13}
  & & \textbf{Min} & \textbf{Max} & \textbf{Avg} & & \textbf{Min} & \textbf{Max} & \textbf{Avg} & & \textbf{Min} & \textbf{Max} & \textbf{Avg} \\
  \midrule

\multirow{2}{*}{\makecell[l]{\textbf{label-}\\\textbf{flipping}}} & \textbf{DDaBA} & 3 & 5 & \textbf{4,92} & & 3 & 5 & 4,87 & & 3 & 5 & 4,92\\
& \textbf{RAB\textsuperscript{2}-DEF} & 3 & 5 & 4,43 & & 3 & 5 & \textbf{4,95} & & 3 & 5 & 4,92\\

\midrule
\multirow{2}{*}{\makecell[l]{\textbf{random}\\\textbf{weights}}} & \textbf{DDaBA} & 3 & 5 & \textbf{4,96} & & 3 & 5 & \textbf{4,96} & & 3 & 5 & 4,96\\
& \textbf{RAB\textsuperscript{2}-DEF} & 3 & 5 & 4,88 & & 3 & 5 & 4,89 & & 3 & 5 & 4,96\\

\midrule
\multirow{2}{*}{\makecell[l]{\textbf{backdoor}\\\textbf{attack}}} & \textbf{DDaBA} & 1 & 5 & \textbf{3,83} & & 1 & 5 & \textbf{3,52} & & 1 & 5 & 3,39\\
& \textbf{RAB\textsuperscript{2}-DEF} & 3 & 5 & 4,65 & & 4 & 5 & 4,35 & & 3 & 5 & 4,18\\
  
\bottomrule
\end{tabular}}
\label{tab:adv}
\end{table*}

\begin{table*}[!h]

\centering
\caption{Minimum (\textbf{Min}), maximum (\textbf{Max}) and average (\textbf{Avg}) number of poor clients discarded by DDaBA and RAB\textsuperscript{2}-DEF throughout the learning rounds for the \textit{label-flipping} and \textit{random weights} attacks.}
\resizebox{\linewidth}{!}{\begin{tabular}{llccccccccccccc}
\toprule
 & & \multicolumn{3}{c}{\textbf{Fed-EMNIST}} & & \multicolumn{3}{c}{\textbf{Fashion MNIST}}  & & \multicolumn{3}{c}{\textbf{CIFAR-10}}\\
 \cmidrule{3-13}
  & & \textbf{Min} & \textbf{Max} & \textbf{Avg} & & \textbf{Min} & \textbf{Max} & \textbf{Avg} & & \textbf{Min} & \textbf{Max} & \textbf{Avg} \\
  \midrule

\multirow{2}{*}{\makecell[l]{\textbf{label-}\\\textbf{flipping}}} & \textbf{DDaBA} & 0 & 5 & 0,93 & & 0 & 5 & 1,18 & & 0 & 4 & 1,03\\
              & \textbf{RAB\textsuperscript{2}-DEF} & 0 & 2 & \textbf{0,12} & & 0 & 3 & \textbf{0,23} & & 0 & 2 & \textbf{0,15}\\

\midrule

\multirow{2}{*}{\makecell[l]{\textbf{random}\\\textbf{weights}}} & \textbf{DDaBA} & 0 & 2 & 0,25 & & 0 & 3 & 0,28 & & 0 & 2 & 0,31\\
          & \textbf{RAB\textsuperscript{2}-DEF} & 0 & 0 & \textbf{0} & & 0 & 1 & \textbf{0,03} & & 0 & 0 & \textbf{0}\\

\midrule

\multirow{2}{*}{\makecell[l]{\textbf{backdoor}\\\textbf{attack}}} & \textbf{DDaBA} & 0 & 3 & 1,20 & & 0 & 3 & 0,98 & & 0 & 4 & 1,72\\
& \textbf{RAB\textsuperscript{2}-DEF} & 0 & 2 & \textbf{0,2} & & 0 & 1 & \textbf{0,25} & & 0 & 2 & \textbf{0,31}\\
  
\bottomrule
\end{tabular}}
\label{tab:poor}
\end{table*}

The filtering statistics of adversarial clients shown in Table \ref{tab:adv} indicate that there are no significant differences between DDaBA and RAB\textsuperscript{2}-DEF. Both approaches effectively filter out all adversarial clients. This observation aligns with the previous performance comparison: both methods perform similarly in filtering adversarial clients. However, when examining the filtering statistics for poor clients in Table \ref{tab:poor}, we notice substantial differences between the two algorithms. When focusing on the maximum number of poor clients filtered (columns labeled \textbf{Max}), DDaBA discards all poor clients in some situations, whereas RAB\textsuperscript{2}-DEF does not discard as many clients in any round. Furthermore, in the average results (columns labeled \textbf{Avg}), DDaBA discards approximately one poor client on average across all rounds in the \textit{label-flipping} attack. RAB\textsuperscript{2}-DEF, however, discards almost no poor clients in all learning rounds. This outcome verifies that RAB\textsuperscript{2}-DEF can differentiate between adversarial and poor clients, discarding only the former. This not only produces better results in some cases (as shown by the previously discussed simulation results), but also ensures a fairer process for the clients, as only those who truly aim to corrupt the learning process are excluded from the aggregation on the server.

Finally, the \textit{random weights} attack is arguably the easiest scenario where adversarial and poor clients can be distinguished, as it is a model poisoning attack rather than a data poisoning attack. In this scenario, the model updates produced by adversarial clients are completely out of distribution, leading to larger differences from the rest of the model updates and simplifying the distinction between poor and adversarial clients.

\rab{\subsection{Analysis on the performance of poor clients}}

\rab{In this section we analyze the impact of the proposal on the performance of poor clients as we state that RAB\textsuperscript{2}-DEF improve the fairness in terms of poor clients' performance. To this end, we analyze the performance of these clients after the learning rounds on the test set. As in the previous section, we compare with DDaBA. In Table \ref{tab:fairness2} we show the mean accuracy in the original task of poor clients for each dataset on each attack scenario. For that metrics, we consider that when a client is discarded, the local model is the one trained locally, but when it participate in the aggregation, the local model is the aggregated model assigned by the server.}

\begin{table}[h!]
\caption{Mean accuracy in the original task of poor clients after the learning rounds.}\label{tab:fairness2}
\centering
\resizebox{\linewidth}{!}{\begin{tabular}{llccc}
\toprule
 &  & \textbf{Fed-EMNIST}  & \textbf{Fashion MNIST}  &  \textbf{CIFAR-10}\\
\toprule
\multirow{2}{*}{\textbf{label-}\textbf{flipping}} & \textbf{DDaBA} & 0.8428 & 0.7653 &  0.7591\\
 & \textbf{RAB\textsuperscript{2}-DEF}  & \textbf{0.9789} & \textbf{0.8841}  & \textbf{0.8797} \\
 \midrule
\multirow{2}{*}{\textbf{random} \textbf{weights}} & \textbf{DDaBA} & 0.8339 & 0.7231 & 0.7169 \\
 & \textbf{RAB\textsuperscript{2}-DEF}  & \textbf{0.9669} & \textbf{0.8801} & \textbf{0.8578}\\
 \midrule
\multirow{2}{*}{{\textbf{backdoor} \textbf{attack}}} & \textbf{DDaBA} & 0.8401 & 0.7341 & 0.7009 \\
 & \textbf{RAB\textsuperscript{2}-DEF}  & \textbf{0.9622} & \textbf{0.8673} &  \textbf{0.8515}\\
\bottomrule
\end{tabular}}
\end{table}

\rab{The results show that the mean performance of poor clients, regardless of the type of the attack, is far higher when they are not discarded. This fact is justified because RAB\textsuperscript{2}-DEF does not exclude poor clients (see Table \ref{tab:poor}), providing them with the opportunity to participate in the global model and thus benefit from the knowledge shared by all clients in the aggregated model. This analysis strongly supports the fairness to poor clients provided by our RAB\textsuperscript{2}-DEF proposal without losing its qualities of defense and robust aggregator for the federated scheme.}

\section{Conclusions}\label{sec:conclusions}


\final{Adversarial attacks pose a significant threat in FL scenarios. Although substantial efforts have been made in the literature, most existing strategies tend to prevent just against one kind of attack, unfairly exclude clients with low-quality local models and fail to provide explanations for the selection or exclusion of clients in the aggregation process.} This work addresses this gap with the proposed RAB\textsuperscript{2}-DEF \rab{a dynamic, explainable and fair to poor clients defense mechanism against byzantine and backdoor attacks in FL}. The results and further analysis show the following:

\begin{itemize}
    \item RAB\textsuperscript{2}-DEF maintains performance in terms of accuracy and attack mitigation compared to other baselines \major{in both byzantine and backdoor attacks scenarios}.
    \item RAB\textsuperscript{2}-DEF is able to dynamically select the clients to filter out being agnostic to the number of adversarial clients and able to adapt to a changing number of them.
    \item RAB\textsuperscript{2}-DEF, which is based on LLEs, provides visual explanations for the filtering out of clients. 
    \item RAB\textsuperscript{2}-DEF is able to distinguish between poor and adversarial clients, ensuring a fair selection of clients \rab{resulting in more robust results both in the global model and the local models of the poor clients}. 
\end{itemize}

In summary, RAB\textsuperscript{2}-DEF ensures robustness, data privacy, integrity and attack mitigation, \rab{and} it also provides other desired requirements for trustworthy AI \rab{stressing explainability and fairness to poor clients.} 


\section*{Acknowledgments}
This research results from the Strategic Project IAFER-Cib (C074/23), as a result of the collaboration agreement signed between the National Institute of Cybersecurity (INCIBE) and the University of Granada. This initiative is carried out within the framework of the Recovery, Transformation and Resilience Plan funds, financed by the European Union (Next Generation).





\bibliography{generalbibfile}

\bibliographystyle{unsrt}


\end{document}